\documentclass[useAMS,usenatbib]{mn2e}
\usepackage{graphicx}
\usepackage{graphics}
\usepackage{epstopdf}
\usepackage{float}
\usepackage{bm}
\usepackage{epsf}
\usepackage{url}

\title[The Universe as a Cellular System]{The Universe as a Cellular System}
\author[Aragon-Calvo M.A. et al.]{M.A. Aragon-Calvo$^{1,2}$\thanks{E-mail:miguel@pha.jhu.edu}\\
$^{1}$Department of Physics and Astronomy. University of California, Riverside, CA, USA.\\
$^{2}$ Department of Physics and Astronomy. Johns Hopkins University. Baltimore, MD 21218, USA.\\}
\begin{document}

\date{}

\pagerange{\pageref{firstpage}--\pageref{lastpage}} \pubyear{2002}
\maketitle
\label{firstpage}

\begin{abstract}

Cellular systems are observed everywhere in nature, from crystal domains in metals, soap froth and cucumber cells to the network of cosmological voids. Surprisingly, despite their disparate scale and origin all cellular systems follow certain scaling laws relating their geometry, topology and dynamics. Using a cosmological N-body simulation we found that the Cosmic Web, the largest known cellular system, follows the same scaling relations seen elsewhere in nature. Our results extend the validity of scaling relations in cellular systems by over 30 orders of magnitude in scale with respect to previous studies. The dynamics of cellular systems can be used to interpret local observations such as the ``local velocity anomaly" as the result of a collapsing void in our cosmic backyard. Moreover, scaling relations depend on the curvature of space, providing an independent measure of geometry.
\end{abstract}

\section{Introduction}

The Cosmic Web is a dynamic system evolving under the action of gravity towards the configuration of minimum energy. Tiny density fluctuations in the primordial density field grew accentuated by gravity to form the galaxies and their associations we observe today. The distribution of galaxies forms an interconnected network of flat walls, elongated filaments and compact clusters of galaxies delineating vast empty void cells. Voids, walls, filaments and clusters can be intuitively associated to cells, faces, edges and vertices in cellular systems
\citep{Einasto80, Ike84, Icke87, Weygaert89,Dubinski93, Weygaert94,Neyrinck14} (see Fig. 1). The geometry and topology of the network of voids has a tantalizing similarity to other cellular systems observed in nature. Even its evolution resembles the coarsening of soap foam where large bubbles grow by the collapse of adjacent smaller bubbles \citep{Plateau73} in a similar way as voids grow by their own expansion and by the collapse of smaller voids in their periphery \citep{Sheth04}. This similarity in structure and dynamics between voids and other cellular systems is the motivation for this study.

In this work we focus on three well-known scaling laws observed in cellular systems. i).- The Lewis law  \citep{Lewis28}, discovered first in cucumber cells, relates the area of a cell and its number of neighbors $n$ (also known as its degree) as  $n = c_1 + c_2 A$, where $A$ is the area of the central cell and $c_1, c_2$ are constants. ii).- The Aboav-Weaire  law \citep{Aboav70, Weaire99} was found while studying cellular domains  in metallic crystals. It states that there is an inverse relation between the degree of a cell and the degree of its adjacent cells. This is usually expressed as  $m_n = a + b/n$, where $m_n$ is the degree of its adjacent cells. iii). The von Neumann law \citep{Neumann52} describing the evolution of two-dimensional foam-cells is expressed in its simplest form as: $dA/dt = k(n-6)$, where $dA/dt$ is the change in the area of the cell and $k$ is a constant. This remarkable relation is purely topological, depending only on the connectivity of cells. The long-sought extension to three dimensions of the von Neumann relation was recently derived by \citet{MacPherson07} (see also \citet{Lazar14} for a recent study of scaling relations on Voronoi systems).

\begin{figure}
  \centering
  \includegraphics[width=0.4\textwidth,angle=0.0]{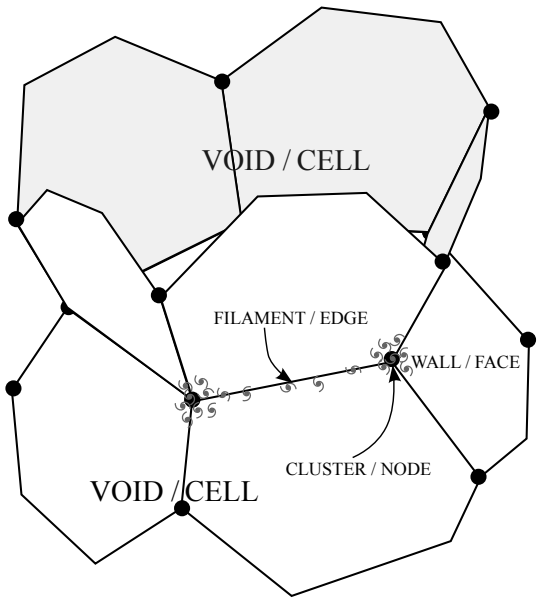}
  \caption{Void cell. Two voids sharing a wall. The voids, walls, filaments and clusters delineated by luminous galaxies correspond to cells, faces, edges and nodes of a cellular system: the Cosmic Foam. For illustration purposes we show galaxies delineating two nodes (clusters) and their common edge (filament).}
  \label{fig:cosmic_web_cellular_system}
\end{figure} 

%
\section{Results}

Our results are based on a dark-matter cosmological N-body simulation with $128^3$ particles inside a box of  $256 h^{-1}$ Mpc side (see Appendix for details). Figure 2 shows three scaling relations (Lewis, Aboav and von Neumann laws) computed from the void network at two different times. We found that, consistent with the Lewis law, large voids have on average a higher degree than smaller voids (Figure 2a).  At the present time the relation is linear while at high redshift and small radius  ($R_{void} < 5$) there is a small departure from linearity.  The peak of the void radius distribution is $R_{void} \sim 7 h^{-1}$ Mpc at $z = 0$ (Figure 3, Appendix) corresponding to $n \sim 14$ in Figure 2a. This value is close to the mean number of neighbors for Voronoi distributions $n_{voro}= 48/35 \pi^2 + 2 = 15.54$  \citep[for an extensive study on Voronoi cells]{Meijering53, Weygaert94}. Predictions for a void network arising from a phase transition give $n = 13.4$ \citep{Laix99} also within our measured values. At early times there is a stronger dependence of $n$ on $R_{void}$ than at the present time. In fact at $z=10$ the mean degree of voids is closer to the Voronoi case. This is consistent (at least qualitatively) with the study by \citet{Weygaert94} who found that the mean degree of Voronoi cells from a Poisson distribution decreases as the Poisson seeds become correlated, this being a crude approximation to the correlating effect of gravity on an initial Poisson distribution of Voronoi seeds.

Figure 2b shows the average neighbor number $m$ of the neighbors cells as function of the number of neighbors $n$ of the central cell (Aboav law). Voids follow the Aboav law at all times. At high redshifts and low $n$ a small decrease in the slope can be observed, although we do not see the full downturn measured in Voronoi distributions \citep{Hilhorst06, Lazar14}. The higher values of $m$ at early times, closer to $n$, suggest a more uniform initial distribution of void sizes.

Figure 2c shows the rate of change in the void size with respect to their degree (von Neumann law). While there is a clear relation at all times, the dispersion increases at low redshift possibly reflecting non-linear processes in the void evolution. There is a small indication that, at very early times voids larger than a couple of Mpc did not collapse. At latter times voids with a lower number of adjacent voids than a critical value $n_{crit}$ will in average collapse. This critical degree is $n_{crit} \sim 18$ at the present time. The Lewis law in addition to von Neumann law implies that below a critical radius $R_{crit} \sim 9 h^{-1}$ Mpc voids collapse and above it they expand (\citet{Sutter14}).

\begin{figure*}
  \centering
  \includegraphics[width=0.99\textwidth,angle=0.0]{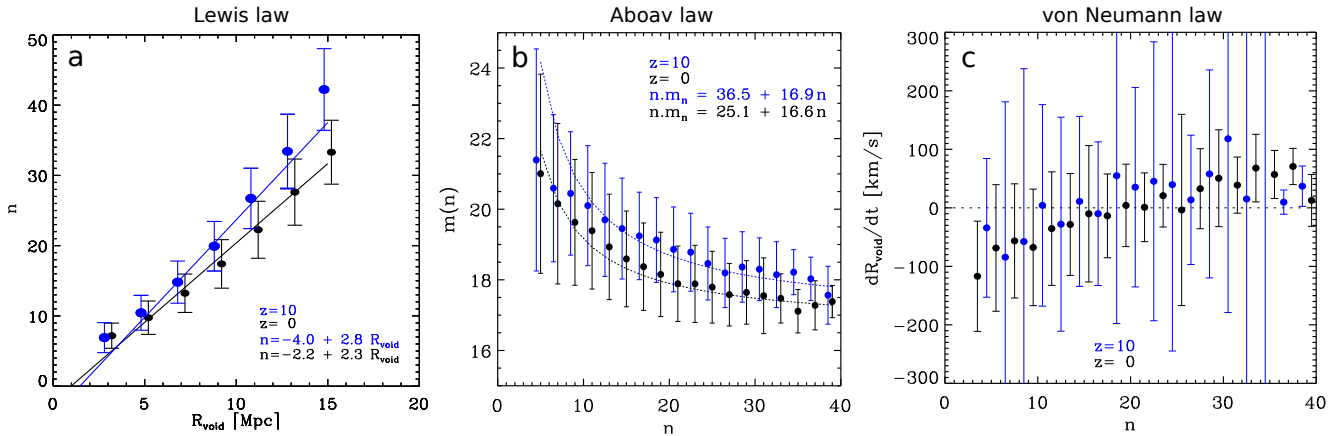}
  \caption{ Scaling relations in Voids, a) Lewis law, b) Aboav law, c) von Neuman Law. (a) Lewis law for the Void system at $z=10$ and $z=0$. $R_{void}$ is the effective radius and $n$ is the number of adjacent voids. The solid line is a linear fit with parameters indicated on the bottom-right. b).- Aboav's law for the Void Network at $z=10$ and $z=0$. The mean inside each bin is indicated by dots, the dotted lines shows a fit to a hyperbolic curve with parameters indicated on the top-right. c).- Von Neumann relation for the void network shown as the rate of change of the void comoving radius $dR_{void}/dt$ as a function of the void's degree. This is the mean velocity of the galaxies at the boundaries of the void measured from the void's center in comoving coordinates. The bars indicate the dispersion inside each bin in all plots.}
  \label{fig:scaling_relations}
\end{figure*}

%
\section{Discussion}

\subsection{A collapsing void in our cosmic backyard}

The geometry and dynamics of our own local cosmic environment has some similarities to a collapsing void scenario. In particular the existence of a population of luminous galaxies off the plane of the local wall \citep{Peebles10}. If our local wall was formed by the collapse of a small void one would expect galaxies at opposite walls of the collapsing void to pass through the newly formed wall \citep{Benitez13}. This scenario is plausible given our position at the edge of a large supercluster in which the Milky Way and its surrounding voids are embedded inside a large shallow overdensity. We in fact have a dramatic example of such collapsing structures in our own cosmic backyard. Just opposite to the local void there is a filament of galaxies, the Leo Spur \citep{Tully08b}, on the farther side of the ``southern void" (the small void opposite to the Local Void) at a distance of $\sim 7$ Mpc approaching to us as a whole with a radial velocity of $\sim 200$ km/$s$. At that distance the velocity discrepancy with the unperturbed Hubble flow is of the order of 700 km/s! This ``local velocity anomaly" \citep{Tully08b} can not be fully explained by the nearby massive structures \citep{Tully08}. It is, however, consistent with the void-in-cloud scenario described by \citet{Sheth04} From figures 2a and 2c we have that $n_{crit}=18$ and $R_{crit} \sim9$ Mpc  comoving. If we add the Hubble expansion factor to convert to an observable velocity then $R_{crit, obs}$ is of the order of $\sim 1$ Mpc which is significantly lower than the radius of the ``southern void" ($\sim 3.5$ Mpc, \citep{Tully08b}). $R_{crit, obs}$ is the mean radius and so we should not be surprised to find variations due to particular LSS configurations. This high value in the rate of collapse of the southern void is worth further study.

%
\subsection{The Void network as a standard ruler}

In general cellular scaling relations assume a Euclidean geometry, however other space metrics are possible. The derivation of the von Neumann law assumes a flat geometry where the integral of the mean curvature $k$ around a cell is $2\pi$. \citet{Avron92} generalized the von Neumann law to curved spaces as $dA/dt = K( (n-6) +  (3A)(\pi R^2))$, where $R$ is the radius of curvature. \citet{Roth12} found a departure from the euclidean von Neumann law in 2D froth on a spherical dome interpreted as a result of the positively curved space being able to accommodate larger angles than the flat space. In the case of the network of Voids we should expect a dependence on the space metric resulting from different geometrical constraints. The Lewis, Aboav and von Neumann laws directly reflect the metric of space by means of the void-connectivity and size and therefore can provide a standard ruler for cosmology. The Lewis and Aboav relations can be measured from galaxy catalogues with accurate galaxy distance estimations and enough spatial sampling to resolve individual voids. In order to measure the von Neumann relation we need actual physical distances (in contrast to redshift distances) which can complicate its measurement. Furthermore, the role of spatial curvature is expected to be marginal. For instance, the two-dimensional von Neumann law has a dependence on spatial curvature as $R^{-2}$. The radius of curvature of the Universe is given by $R = D_0 / \sqrt{\Omega -1}$, where  $D_0 = H_0^{-1} \simeq 3000 h^{-1}$Mpc and $\Omega$ is the total density of the Universe. Any effect of spatial curvature on the von Neumann law will be very small for reasonable values of $\Omega$. The simulations we present here can not account for non-euclidian geometries. More complete simulations, better theoretical understanding and future massive 3D galaxy surveys with accurate distance estimates are required to apply this relations as an independent standard ruler for cosmology.

\section{Acknowledgements}
This research was partly funded by the Betty and Gordon Moore foundation and by a New Frontiers of Astronomy and Cosmology grant from the Templeton Foundation. The author would like to thank Mark Neyrinck for stimulating discussions.

\appendix 

\section{Computer simulations and void catalogues}

The results presented in this work are based on a N-body computer simulation containing $128^3$ dark matter particles inside a box of $256$ h$^{-1}$ side with the standard $\Lambda$CDM cosmology with $\Omega_m=0.3$, $\Omega_{\Lambda}=0.7$, $h=0.73$, $\sigma_8=0.8$. Starting at $z=80$ we evolved the box to the present time using the N-body code {\small GADGET-2} \citep{Springel05} and stored 32 snapshots in logarithmic intervals of the expansion factor starting at $z=10$. While the particle number may seem too low in fact for LSS studies it is sufficient since the smallest voids we are interested in are larger than a few Mpc in radius. The low particle number imposes a low-pass filter, removing unwanted structures arising from over-segmentation in the watershed method used to identify voids.

\subsection{Void identification and tracking}
\begin{figure}
  \centering
  \includegraphics[width=0.4\textwidth,angle=0.0]{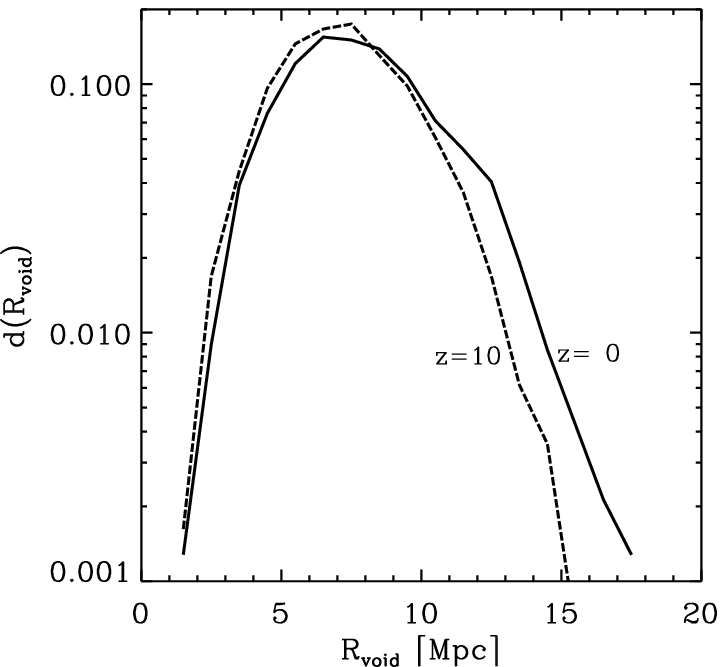}
  \caption{Distribution of void sizes $R_{\textrm{\tiny{void}}}$ at $z=0$ (solid line) and $z=10$ (dashed line). }
  \label{fig:radius_distribution}
\end{figure}

\begin{figure}
  \centering
  \includegraphics[width=0.4\textwidth,angle=0.0]{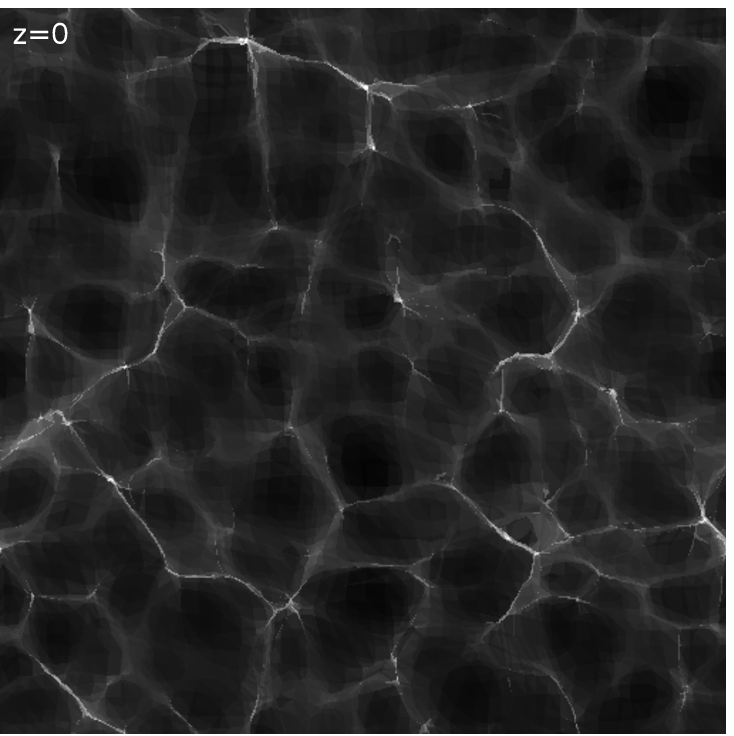}
  \caption{Density field computed with the Lagrangian Sheet approach $z=0$. This density estimation is optimal for the near-regular particle distribution in the under-dense regions inside voids}
  \label{fig:density fields}
\end{figure}

From the particle distribution at each of the 32 snapshots we computed a continuous density field on a regular grid of $512^3$ voxel size using a Lagrangian Sheet approach \citep{Abel12, Shandarin12} as described in \citet{Aragon14}. The Lagrangian nature of this density estimation and interpolation method allows us to compute accurate densities at very early times and also at latter times inside voids where the particle arrangement is still close to a regular grid (see Fig \ref{fig:density fields}). Next we identified voids using the floating-point implementation of the watershed transform in the Spine pipeline \citep{Aragon10a}. In order to further minimize over-segmentation of voids arising from spurious splitting of voids between snapshots we merged voids in adjacent snapshots if a void at  given snapshot had more than $\% 70$ of its volume in the next snapshot. The distribution of void sizes at two times is shown in Fig. \ref{fig:radius_distribution}. The effective void radius was computed from its volume as $R_{\textrm{\tiny{void}}} = ( (3/4\pi) V_{\textrm{\tiny{void}}} )^{1/3}$. For each void we identified its adjacent voids (voids that share a common wall) and created a \textit{void-graph} with nodes corresponding to void centers and edges joining adjacent voids (this \textit{void-graph}  is a triangulation and is the dual of the \textit{cosmic web-graph} which is a cellular system). This void-graph was used to compute the Lewis and Aboav relations in Fig. 2. In order to trace the evolution of individual voids (for the von Neumann relation) we created a \textit{void progenitor line} in a similar way as in done by \citet{Aragon10b} but performing the linking across time (adjacent snapshots) instead of scale (hierarchical space).

\end{document}